\documentclass[twocolumn,preprint]{aastex62}

\usepackage[T1]{fontenc}
\usepackage[english]{babel}
\usepackage{natbib}
\usepackage{ulem,xcolor}
\usepackage{amsmath, graphicx}
\usepackage{soul}
\begin{document}


\title{Pre-impulsive and Impulsive Phases of the Sub-Terahertz Flare of March 28, 2022}%

\author[0000-0001-7856-084X]{G.G. Motorina}
	\affil{Central  Astronomical Observatory at Pulkovo of Russian Academy of Sciences, St. Petersburg, 196140, Russia}

\affil{Ioffe Institute, Polytekhnicheskaya, 26, St. Petersburg, 194021 Russia}

\affil{Astronomical Institute of the Czech Academy of Sciences, 251 65 Ond\v{r}ejov, Czech Republic}

\author[0000-0001-5074-7514]{Yu.T. Tsap}
	\affil{Crimean Astrophysical Observatory, Russian Academy of Sciences, Nauchnyi, Russia}

\author[0000-0002-4018-847X]{V.V. Smirnova}
	\affil{Crimean Astrophysical Observatory, Russian Academy of Sciences, Nauchnyi, Russia}

\author{A.S. Morgachev}
	\affil{Central Astronomical Observatory at Pulkovo of Russian Academy of Sciences, St. Petersburg, 196140, Russia}

\author{A.D. Shramko}
	\affil{Central Astronomical Observatory at Pulkovo of Russian Academy of Sciences, St. Petersburg, 196140, Russia}

\author{A.S. Motorin}
	\affil{ITMO University, 197101 St. Petersburg, Russia}

\received{February 27, 2023 }

\begin{abstract}
Properties of the solar radio spectrum, as well as the temporal profiles of flare emission, indicate the thermal nature of the sub-terahertz (sub-THz) component observed as the growth of radio emission in the frequency range of 100–1000 GHz. 
The sub-THz flare onset can be ahead of the impulsive phase for several minutes. However, the origin of the pre-impulsive and impulsive sub-THz emission remains unclear. The
present work is devoted to a detailed analysis of the M4.0 X-class solar flare observed on March 28, 2022 with
the Bauman Moscow State Technical University Radio Telescope RT-7.5 at 93 GHz. We supply these data with
multiwavelength solar observations in the X-ray (GOES, GBM/Fermi), extreme ultraviolet (AIA/SDO), and
microwave ranges. The differential emission measure (DEM) responsible for EUV emission is determined by
solving the inverse problem based on the AIA/SDO data. Using the DEM and assuming a thermal free-free
emission mechanism in pre-impulsive and impulsive phases, we calculated the millimeter emission flux of coronal
plasma of the flare source, which turned out to be much smaller than the observed values. We concluded
that electrons accelerated in the corona and heat fluxes from the coronal loop top cannot be responsible for heating
the sub-THz emission source located in the transition region and upper chromosphere. A possible origin of
chromospheric heating in the pre-impulsive phase of the solar flare is discussed.
\end{abstract}



\section{Introduction}
\label{S_Intro}
Relatively recently, it became possible to carry out regular solar observations in the sub-terahertz (sub-THz) range of the electromagnetic spectrum at frequencies
of 100-1000 GHz \citep{2001ApJ...548L..95K, 2013A&ARv..21...58K}. 
This has opened a new window of opportunity to study the nature of solar flares. In particular,
the unusual increase in sub-THz flux between 212 and 405 GHz during the solar flare on March 22, 2000 \citep{2001ApJ...548L..95K}, the mechanism of which is still highly controversial \citep{2013A&ARv..21...58K, 2018A&A...620A..95K, 2018SoPh..293...50T}, marked the emergence
of a new direction in solar radio astronomy research.

Modern Atacama Large Millimeter Array (ALMA)  observations of the Sun  \citep{2016SSRv..200....1W, 2017A&A...605A..78A} have allowed to obtain the structure of the upper layers of the solar atmosphere
in the previously inaccessible sub-THz range. As of today, only six weak solar flares have been registered with ALMA, one event \citep{2021ApJ...922..113S} in
the interferometric mode at 100 GHz and five flares \citep{2023A&A...669A.156S} at 100 GHz ($\lambda = 3$ mm) or 240 GHz ($\lambda = 1.2$ mm) in the single-dish scanning mode, as
necessary for image calibration. It has been shown (see, e.g., \cite{2017A&A...605A..78A})
that images of the transition region and upper chromosphere observed in
the sub-THz and ultraviolet (1600 \AA) ranges exhibit similar magnetic structures. In contrast, based on observations at 1.2 and 3 mm with spatial resolutions of about $28''$ and $60''$  respectively, \cite{2023A&A...669A.156S} identified sources of millimeter radiation with hot loops visible in SDO/AIA. However, the temporal resolution of these observations was low, approximately
$\sim5$ (3 mm) and $\sim10$ (1.2 mm) minutes, making the
conclusions of \cite{2023A&A...669A.156S} less convincing.

The problem of flare atmospheric heating and energy transfer of accelerated particles to the dense
layers of the chromosphere/photosphere remains unresolved. In particular, X-ray sources in the preimpulsive phase of a flare may be located at the footpoints of coronal loops, and their temperatures may reach 10-15 MK  \citep{2021MNRAS.501.1273H}. On the other hand, sub-THz emission is associated with X-ray emission \citep{2009ApJ...697..420K}, for the generation
of which the thermal plasma of the corona, transition region, and upper chromosphere may be responsible. This suggests that the study of the sub-THz and X-ray sources can give valuable information about the heating of the chromospheric plasma, which can be caused
by accelerated electrons, thermal fluxes, shock waves, as well as \textit{in situ} flare energy release.

The aim of this study is to investigate the role of coronal thermal plasma and accelerated electrons in generating the sub-THz radiation based on the SOL2022-03-28T105800 flare event, as well as to try to understand the nature of heating in the transition region and upper chromosphere of the Sun, primarily in the pre-impulsive phase of the solar flare.

\section{OBSERVATIONAL DATA AND PROCESSING}

The SOL2022-03-28T105800 solar flare was selected for the analysis of the heating source in the preimpulsive phase, since it was recorded by the RT-7.5
radio telescope of the Bauman Moscow State Technical University (93 GHz) and was observed by several other instruments across a wide wavelength range. The
SOL2022-03-28T105800 flare of X-ray class M4.0 occurred in active region NOAA 12975 between 10:58 and 11:45 UT, with the maximum of sub-THz emission at 11:28 UT.

\begin{figure*}\centering
\includegraphics[width=8cm]{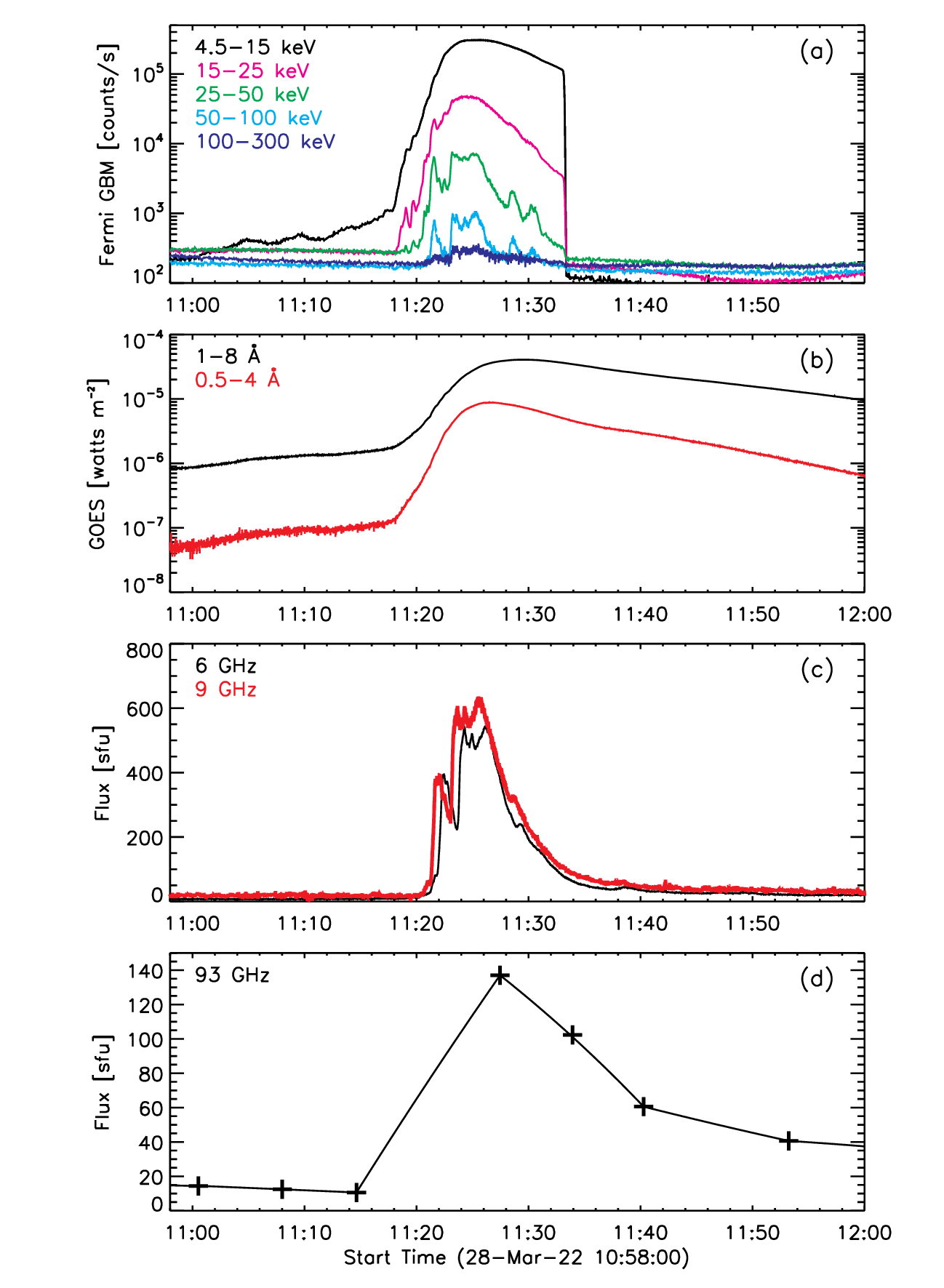}
\includegraphics[width=6cm]{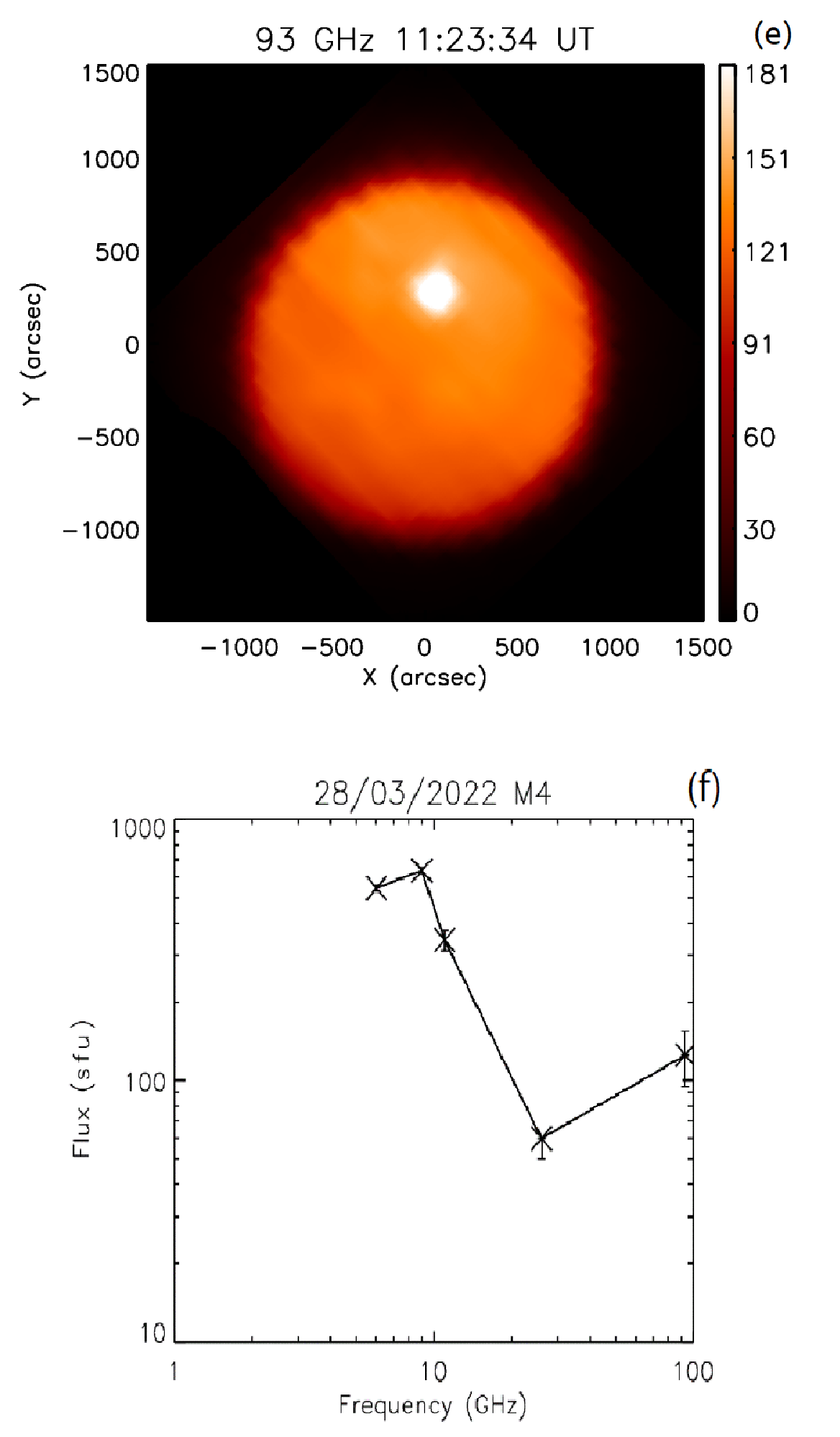}
\caption{Left: the time profiles of the SOL2022-03-28T105800 solar flare in the X-ray (a, b), microwave (c), and sub-THz (d) ranges.
Right: the flare map at 93 GHz (e), the flare flux density spectrum (f).}    
\label{Fig1}
\end{figure*}

To accomplish the task, the extreme ultraviolet (EUV) data at 94, 131, 171, 193, 211, and 335 \AA \;wavelengths were analyzed. These data were obtained using the Atmospheric Imaging Assembly on board Solar Dynamic Observatory (AIA/SDO) \citep{2012SoPh..275...17L}. The X-ray radiation data from the Fermi Gammaray Burst Monitor (GBM/Fermi) \citep{2009ApJ...702..791M} and the Geostationary Operational Environmental
Satellite (GOES) \citep{2005SoPh..227..231W} were also
employed (Figs.\ref{Fig1}a, \ref{Fig1}b). 
These instruments provide information about the thermal flare plasma and nonthermal
accelerated electrons. Furthermore, for a more detailed analysis of the SOL2022-03-28T105800 flare the data from the Kislovodsk Mountain Astronomical
Station obtained with RT-2 and RT-3 radio telescopes
at frequencies of 6 and 9 GHz, respectively 
(\url{http://solarstation.ru/}, \cite{Shramko}), the RT-7.5 radio
telescope of Bauman Moscow State Technical University (93 GHz, \cite{1981IzVUZ..24....3R};  \cite{Smirnova}) (Figs.\ref{Fig1}c, \ref{Fig1}d), 
Metsähovi Radio Observatory
(11.2 GHz, \cite{2018AN....339..204K}), and the observatory of
the Italian National Institute for Astrophysics (INAF,
26 GHz, \cite{2022SoPh..297...86P}) were included.

The time profile of sub-THz radiation at 93 GHz
(Fig.\ref{Fig1}d) was constructed from the maps (Fig.\ref{Fig1}e)
obtained with the RT-7.5 radio telescope. In each
map, the active region in which the flare occurred was
identified, followed by the calculation of the maximum
flux value. It is evident from Fig.\ref{Fig1}d that during the pre-impulsive phase of the flare from 11:00:00 to
11:00:12 UT, the spectral flux density of sub-THz radiation
increases, reaching approximately 20 sfu. The
procedures for calibrating the sub-THz radiation flux
at the RT-7.5 radio telescope, including noise estimates
of the receiving equipment, atmospheric
absorption coefficients, tracking error uncertainties,
etc., align with those described by \cite{2016AdSpR..57.1449T} and \cite{Ryzhov}. Thus, we used standard observational calibration methods utilized in sub-THz
(millimeter) radio astronomy \citep{2001ApJ...548L..95K}.

What draws attention first of all is the similarity of
the time profiles of microwave and hard X-ray
emission (Figs.\ref{Fig1}a, \ref{Fig1}c), which consisted of multiple
peaks. Although observations in the sub-THz range
were carried out only at 93 GHz, the spectral flux density
spectrum of the SOL2022-03-28T105800 solar
flare (Fig.\ref{Fig1}f) indicates that it was positive between
frequencies of 26 and 93 GHz, with a power-law index
$\alpha \approx 0.6$. The error estimates are indicated on the spectrum
by bars in Fig.\ref{Fig1}f.

\begin{figure*}\centering
\includegraphics[width=12cm]{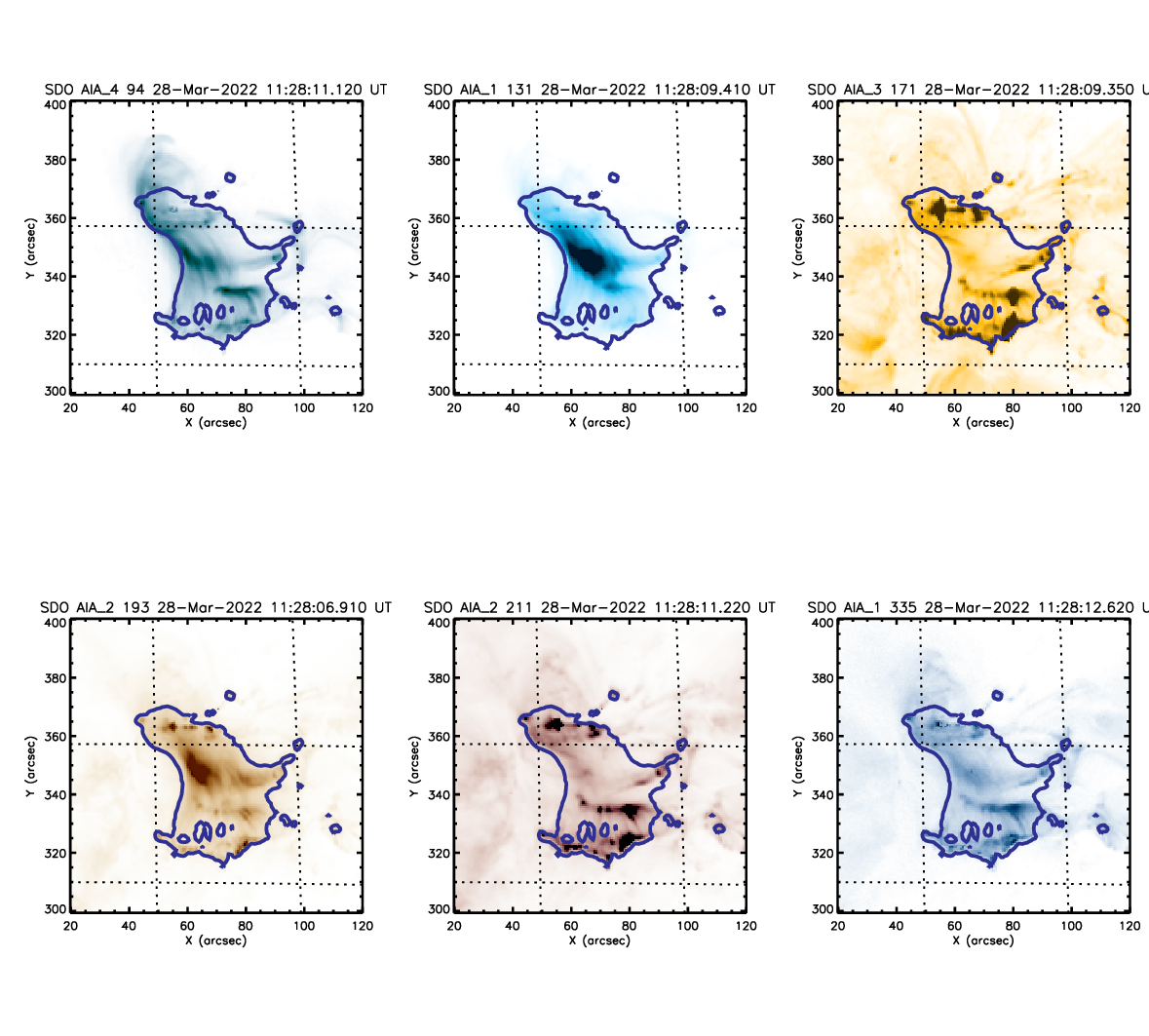}
\caption{The EUV images of the SOL2022-03-28T105800 solar flare obtained at 94, 131, 171, 193, 211, and 335 Å AIA/SDO passbands. The blue line marks the 50\% contour of the intensity maximum.}    
\label{Fig2}
\end{figure*}

\begin{figure*}\centering
\includegraphics[width=7cm]{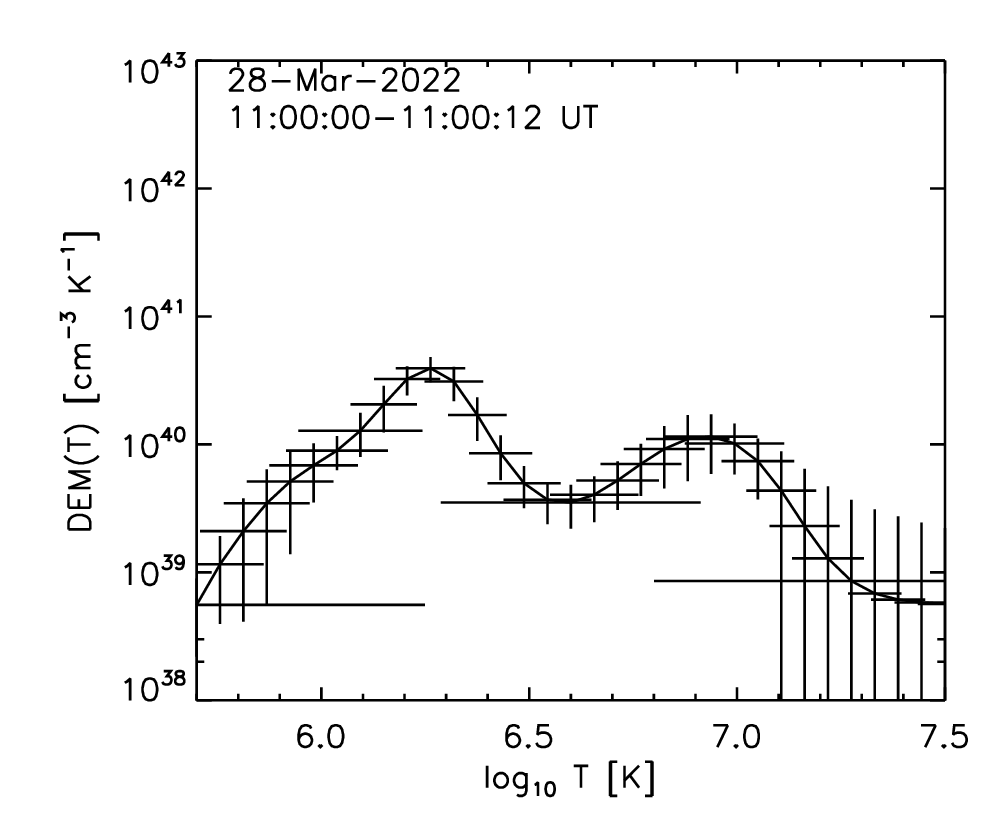}
\includegraphics[width=7cm]{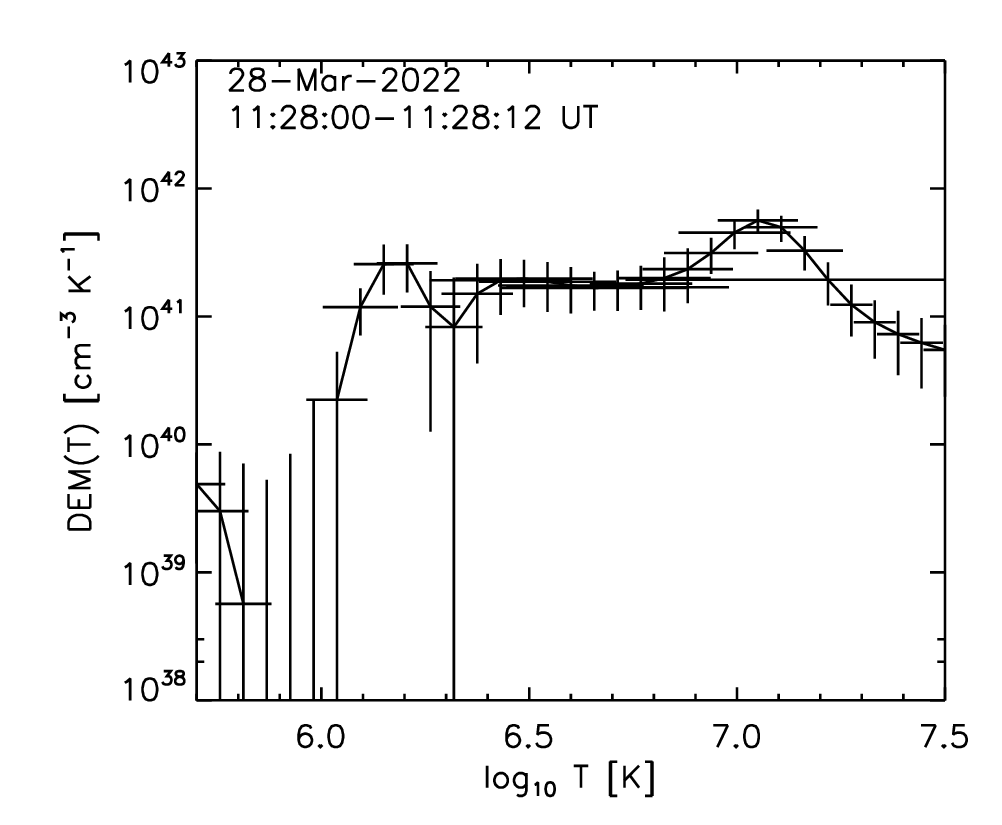}
\caption{The differential emission measure of the SOL2022-03-28T105800 solar flare estimated for the times 11:00:00–11:00:12 UT (left) and 11:28:00–11:28:12 UT (right).}    
\label{Fig3}
\end{figure*}

Using the AIA/SDO data (Fig.\ref{Fig2}) at two time
intervals corresponding to the begining of the flare (11:00:00–
11:00:12 UT) and the peak of the inpulsive phase (11:28:00–11:28:12 UT)
of the millimeter burst, we performed the reconstruction
of the differential emission measure (DEM, Fig.\ref{Fig3})
using the regularization technique \citep{Tikh...1979, 2012A&A...539A.146H}:

\begin{equation}
\phi(T)=n_e^2\frac{dl}{dT},
\end{equation}
where $n_e$ is the electron number density and $l$ is the
source size along the line of sight for flare plasma temperatures
in the range of $T = 0.5-32$ MK.

Based on the obtained distributions $\phi(T)$, calculations
were performed for the millimeter emission
using the integral brightness temperature $T_b$ and the
observed spectral flux $F_{\nu}$ in the assumption of thermal
bremsstrahlung emission according to formulas \citep{1985ARA&A..23..169D, 2016AdSpR..57.1449T}:

\begin{gather}
T_b(\nu)=\frac{1}{\nu^2} \int_{T_{\min}}^{T_{\max}} \frac{K\phi(T)}{\sqrt{T}} e^{-\tau_{\nu}(T)}dT, 
\\
\tau_{\nu}=\int_{T_{\min}}^{T_{\max}} \frac{K\phi(T)}{T^{3/2} \nu^2} dT,
\\
F_{\nu}= \frac{2k_{\rm{B}} \nu^2}{c^2} T_b(\nu)\frac{S}{R^2},
\end{gather}

in which the standard notation is used.

As follows from Eqs. (2-4), even for the maximum
thermal source areas $S = 7.8 \times 10^{18}$ cm$^{-3}$, corresponding
to the 50\% level of the maximum intensity in the
EUV maps (Fig.\ref{Fig2}), the calculated fluxes $F_{\nu}$ at 93 GHz
were 1.5 sfu (11:00 UT) and 45 sfu (11:28 UT), respectively. 
These values were noticeably underestimated
compared to the observed values of $20 \pm 5$ sfu and
$125 \pm 31$ sfu, respectively. Taking into account that filaments in the energy release region
were not visible and the contribution of optically thin
flare plasma with temperatures $T = 0.5-32$ MK to
millimeter emission was negligibly small, as indicated
by our estimates, we can conclude that the sub-THz
component was most likely generated by cool chromospheric
plasma.

\section{SOURCE OF CHROMOSPHERIC PLASMA HEATING}

\begin{figure*}\centering
\includegraphics[width=7cm]{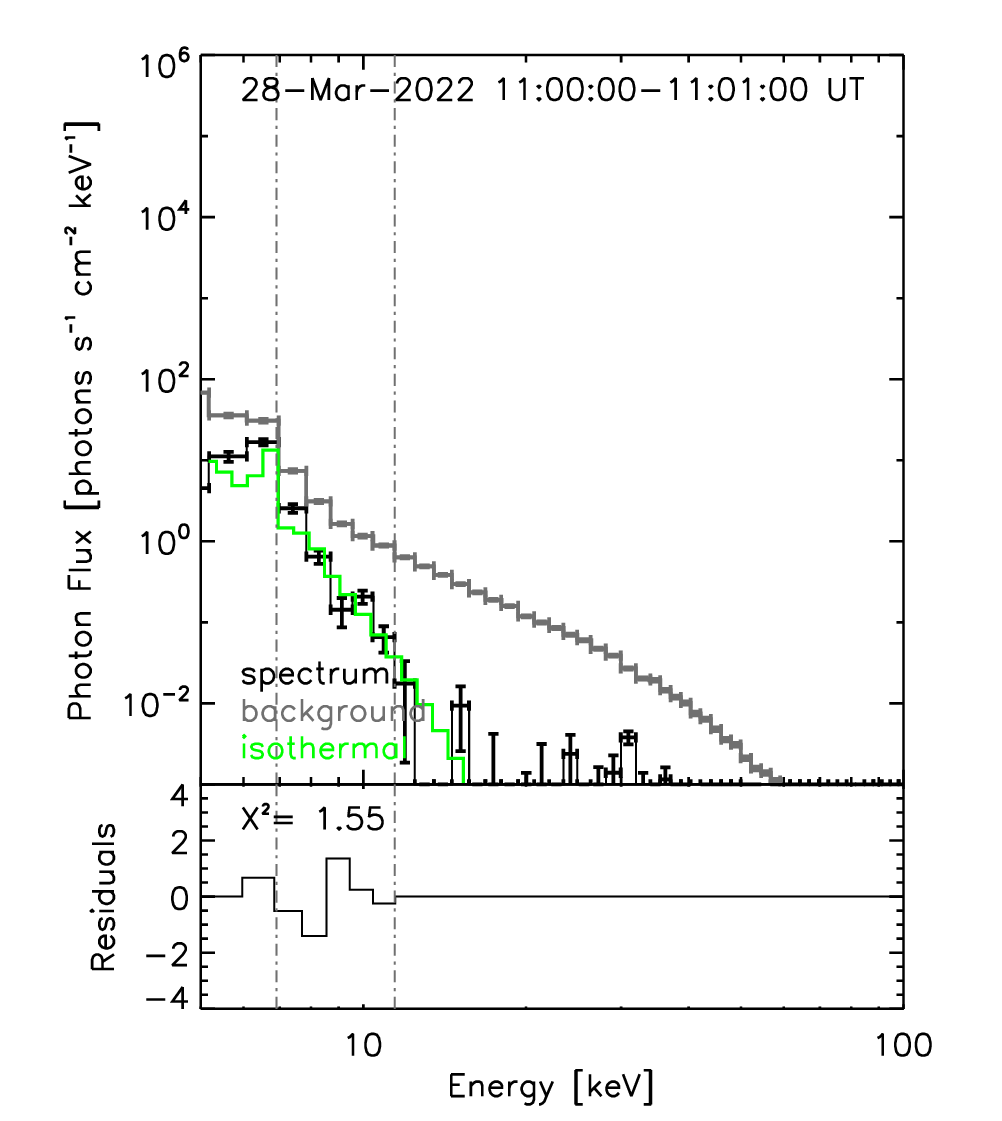}
\includegraphics[width=7cm]{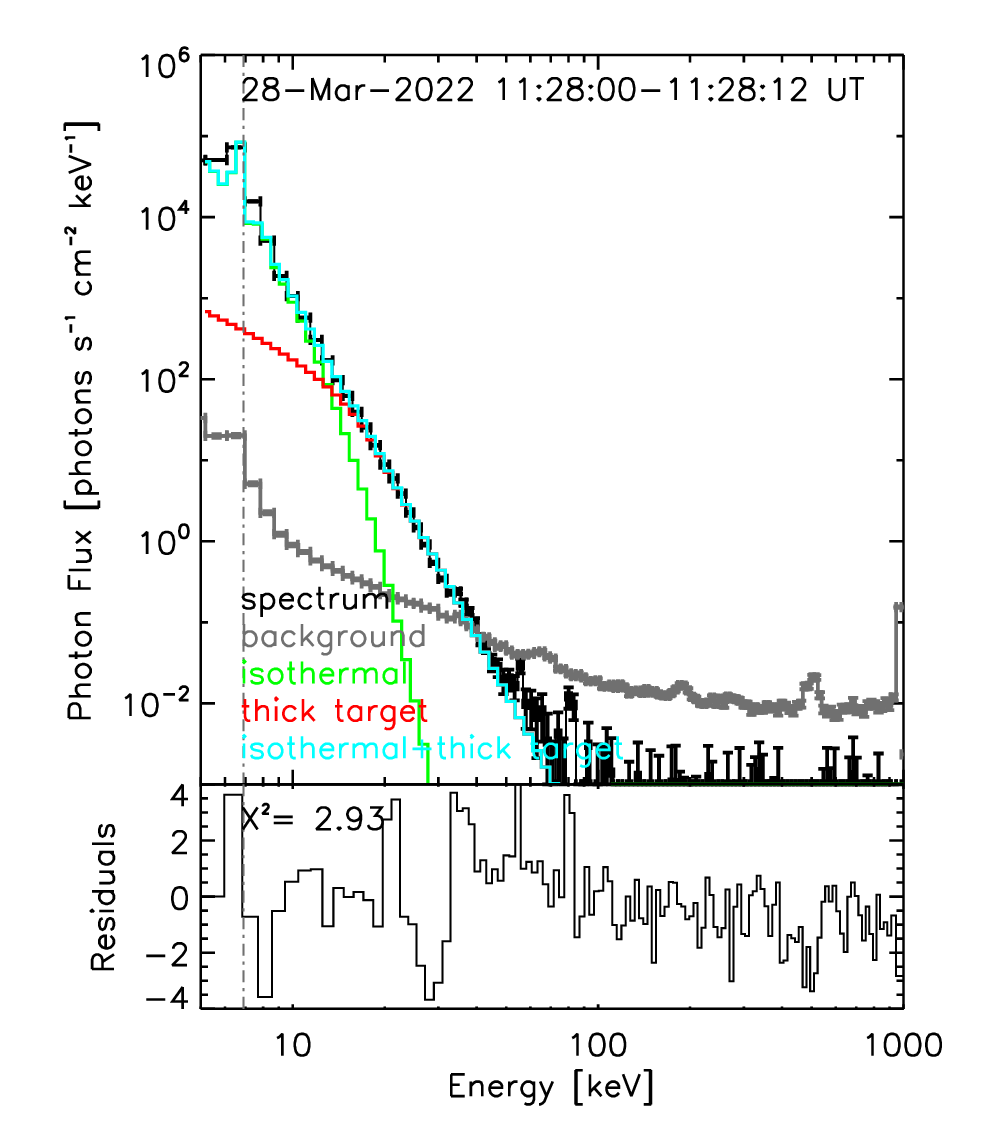}
\caption{The averaged GBM/Fermi X-ray spectrum (background subtracted data, black line) fitted by the homogeneous isothermal plasma (green line) for the time 11:00:00–11:00:12 UT (right); the homogeneous isothermal plasma (green line) and thick target
model (red line) for 11:28:00–11:28:12 UT (left). The summed fit of the spectrum is shown by the blue line. The background is shown by the gray line. The residuals are shown in the bottom panels.}    
\label{Fig4}
\end{figure*}

To estimate the contribution of accelerated electrons
to the observed sub-THz emission of the
SOL2022-03-28T105800 solar flare, we performed
the fitting of the X-ray spectrum using data from the
GBM/Fermi satellite, assuming a homogeneous isothermal
plasma and a thick-target model. In the initial
phase of the flare, 11:00:00–11:00:12 UT, the X-ray
emission was comparable to the background values, so
the fitting of the GBM/Fermi data was done assuming
a homogeneous isothermal plasma excluding the contribution
of nonthemal component (Fig.\ref{Fig4}, left). However,
during the period from 11:28:00 to 11:28:12 UT,
which corresponded to the peak of the sub-THz emission
(Fig.\ref{Fig4}, right), we also considered the nonthermal
component.

This allowed us to estimate within the
thick-target model the spectral index of nonthermal
accelerated electrons $\delta = 7.86$, the low-energy cutoff 
$E_0 = 18.57$ [keV], and the total integrated electron flux 
$F =8.57 \times 10^{35}$ [s$^{-1}$]. As indicated by the obtained results, the temperature of the X-ray plasma in the pre-impulsive
phase of the flare reached $T = 15.37$ MK at negligible
value of emission measure $EM = 10^{45}$ cm$^{-3}$,
which is in a good agreement with the findings of \cite{2021MNRAS.501.1273H}. 
Meanwhile, during the peak,
$EM = 3.89 \times 10^{49}$ cm$^{-3}$ and $T = 17.05$ MK, the
spectrum of accelerated electrons occurred to be too
soft ($\delta = 7.86$) to effectively heat the plasma of the
transition region and chromosphere \citep{Tsap, 2021Ge&Ae..61.1045M}.

Considering the relatively high values of the X-ray
plasma temperature both in the pre-impulsive phase
and the flare peak, it might seem that coronal thermal 
fluxes could be responsible for heating the chromospheric
plasma. Let us make some simple estimates.

The rate of heating of the transition region and
upper chromosphere plasma due to electron thermal
conductivity can be estimated as follows:

\begin{gather}
q=\kappa \frac{d}{ds} (T^{5/2}\frac{dT}{ds}) =\kappa \frac{2}{7} \frac{d}{ds}(\frac{dT^{7/2}}{ds}) 
=\kappa \frac{2}{7} \frac{d^2 T^{7/2}}{ds^2} \approx \kappa \frac{2}{7} \frac{T^{7/2}}{s^2}   
\end{gather}

where the electron thermal conductivity coefficient
$\kappa = 10^{-6}$ erg K$^{-7/2}$ s$^{-1}$. Then, assuming thermal energy
$W_{\rm{th}} = 3/2nkT$, where $k = 1.38 \times 10^{-16}$ [erg K$^{-1}$] is the
Boltzmann constant, plasma density $n = 10^{11}$ [cm$^{-3}$],
the temperature $T = 10^5$ [K], and plasma layer thickness
$s = 3 \times 10^7$ [cm], we obtain the characteristic
heating time of the transition region:

\begin{equation}
    \tau_{\rm{con}}=\frac{W_{\rm{th}}}{q}=21\frac{kns^2}{\kappa T^{5/2}} \approx 10^5 \rm{s}.
\end{equation}

The obtained estimate ($\sim 28$ hours) excludes the possibility
of effective heating of not only the chromosphere
but also the transition region by coronal thermal fluxes
during the flare energy release.

As for shock waves, their significant contribution to
chromospheric heating appears doubtful, especially
since we did not find any indications of explosive
energy release in the pre-impulsive phase.

\section{DISCUSSION OF RESULTS AND CONCLUSIONS}

In this study, we performed an analysis of the sub-
THz, microwave, EUV, and X-ray emission of the SOL2022-03-28T105800 solar flare. This allowed us
to determine the contribution of the coronal flare
plasma and accelerated electrons to the generation of
sub-THz emission in the pre-impulsive and impulsive
(close to the maximum of the flare) phases. It
has been shown that the coronal plasma contributes
insignificantly to the observed millimeter flare emission
at 93 GHz. The parameters of nonthermal electrons,
namely, the soft power-law spectrum and the
absence of the observable hard X-ray and microwave
emission during the initial flare phase indicate the
minor contribution of accelerated electrons to chromospheric plasma heating.

The calculated sub-THz fluxes are an order of magnitude
smaller than the observed values. This suggests
that the contribution of the thermal flare coronal
plasma to the sub-THz component is negligibly small.
Hence, the conclusion about the important role of an
additional energy release source located directly in the
solar chromosphere can be drawn. We believe that there
was \textit{in situ} heating of the flare chromosphere occurred
due to the Joule dissipation of electric current. One of
the possible mechanisms based on the interaction of
ions with neutral atoms in nonstationary conditions is
described in \citep{2016Ge&Ae..56..952S}. 
Further more
detailed investigations require multi-frequency sub-
THz observations with high spatial and temporal resolution, since it cannot be fully excluded that heating
mechanisms of nonthermal nature also played an
important role in the considered case.

\section*{ACKNOWLEDGMENTS}
The study was in part supported by the State Assignment
no. 0040-2019-0025 (G.G. Motorina), Russian Science
Foundation grant no. 22-12-00308 (Yu.T. Tsap, A.S. Morgachev),
the Russian Foundation for Basic Research grant
no. 20-52-26006 (V.V. Smirnova), Ministry of Education
and Science (NIR No. 1021051101548-7-1.3.8; 075-03-
2022-119/1) (A.D. Shramko).

\bibliography{apssamp}

\end{document}